\documentclass[conference]{IEEEtran}
\IEEEoverridecommandlockouts
\usepackage{cite}
\usepackage{amsmath,amssymb,amsfonts}
\usepackage{amsmath, bm}
\usepackage{algorithmic}
\usepackage{algorithm}
\usepackage{graphicx}
\usepackage{textcomp}
\usepackage{xcolor}
\usepackage{array}
\def\BibTeX{{\rm B\kern-.05em{\sc i\kern-.025em b}\kern-.08em
    T\kern-.1667em\lower.7ex\hbox{E}\kern-.125emX}}

\begin{document}
\title{
Resource Allocation for Capacity Optimization in Joint Source-Channel Coding Systems

}

\author{\IEEEauthorblockN{Kaiyi Chi $^{*}$, Qianqian Yang $^{*}$, Zhaohui Yang $^{*}$, Yiping Duan $^{\dagger}$, Zhaoyang Zhang $^{*}$}
\IEEEauthorblockA{
$^{*}$ College of Information Science and Electronic Engineering, Zhejiang University, Hangzhou, China \\
$^{\dagger}$ Department of Electronic Engineering, Tsinghua University, Beijing, China \\
E-mail: \{kaiyichi, qianqianyang20, yang\_zhaohui, ning\_ming\}@zju.edu.cn} yipingduan@mail.tsinghua.edu.cn
}

\maketitle
\begin{abstract}
Benefited from the advances of deep learning (DL) techniques, deep joint source-channel coding (JSCC) has shown its great potential to improve the performance of wireless transmission. However, most of the existing works focus on the DL-based transceiver design of the JSCC model, while ignoring the resource allocation problem in wireless systems. In this paper, we consider a downlink resource allocation problem, where a base station (BS) jointly optimizes the compression ratio (CR) and power allocation as well as resource block (RB) assignment of each user according to the latency and performance constraints to maximize the number of users that successfully receive their requested content with desired quality. To solve this problem, we first decompose it into two subproblems without loss of optimality. The first subproblem is to minimize the required transmission power for each user under given RB allocation. We derive the closed-form expression of the optimal transmit power by searching the maximum feasible compression ratio. The second one aims at maximizing the number of supported users through optimal user-RB pairing, which we solve by utilizing bisection search as well as Karmarkar's algorithm. Simulation results validate the effectiveness of the proposed resource allocation method in terms of the number of satisfied users with given resources.  

\end{abstract}


\section{Introduction}
Traditional communication systems are designed based on Shannon's separation theorem, where source coding and channel coding are two separate steps. The former is used to remove the redundancy of the source data, while the latter is used to add redundant information to enhance the robustness towards noisy channel. It has been proven in Shannon's theory that this two-step approach is optimal theoretically when the size of the transmitted message goes to infinity in memory-less channels \cite{shannon}. Nevertheless, in practical applications we are limited to finite block length, and previous research has shown that joint source-channel coding outperforms the separate scheme in the finite block length regime \cite{joint}, motivating the design of joint source-channel coding schemes. Nowadays, with the development of deep learning, some JSCC models have been used in text transmission \cite{deepsc}, image transmission \cite{jscc}, \cite{multi-level}, speech transmission \cite{speech}, etc., which have shown remarkable advantages over traditional solutions.

Some efforts have been made towards the resource allocation problem in JSCC systems in the literature. \cite{performance} considered a JSCC system for downlink text transmission, where a predefined metric of semantic similarity (MSS) was optimized by jointly optimizing the resource block selection as well as the semantic information to be transmitted. The authors in \cite{adaptive} investigated the resource allocation problem of a latency-sensitive system, in which the edge server performs inference tasks according to the transmitted data of the edge devices. 
\cite{resource} also considered the uplink transmission of a JSCC scheme for text transmission, where the authors first defined 
the semantic spectral efficiency (S-SE), and then maximized the overall S-SE by optimizing the channel assignment and the number of transmitted symbols. 
The aforementioned works mainly focus on maximizing the weighted performance of all users.
However, the limited resource at the BS is not always sufficient to support all users within its coverage. To tackle this issue, we are motivated to propose a novel resource allocation method aiming at maximizing the number of users the BS can serve.

In this paper, we consider the donwlink resource allocation of a JSCC based communication system to serve as many users as possible while taking both latency constraints and performance requirements into account. Specifically, we first define a binary utility function, which reflects whether the latency and performance constraints of the user are met or not. Then we define an optimization problem to maximize this utility function by selecting the compression ratios of the transmitted symbols, and allocating resource blocks and power. To solve the problem, we decouple it into two subproblems, which we have proven it is without lose of optimality. The first subproblem is to minimize the required power of each user-RB pair with the given RBs assignment. Meanwhile, the second subproblem aims to maximize the overall capacity of the system with the power constraint. We solve these two subproblems separately with polynomial-complexity by transforming the user-RB pairing problem into a linear optimization problem and solve it with Karmarkar's algorithm. Simulation results demonstrate the effectiveness of the proposed method over benchmarks in terms of the number of served users.

The rest of this article is organized as follows. Section \uppercase\expandafter{\romannumeral2} introduces the system model. The resource allocation problem is defined in Section \uppercase\expandafter{\romannumeral3}. In Section \uppercase\expandafter{\romannumeral4}, we decompose the original problem into two subproblems, and then we solve these two subproblems separately. 
Section \uppercase\expandafter{\romannumeral5} presents the simulation results. Section \uppercase\expandafter{\romannumeral6} concludes this paper.


\section{System Model}
\begin{figure}[htbp]
\centerline{\includegraphics[width=7cm]{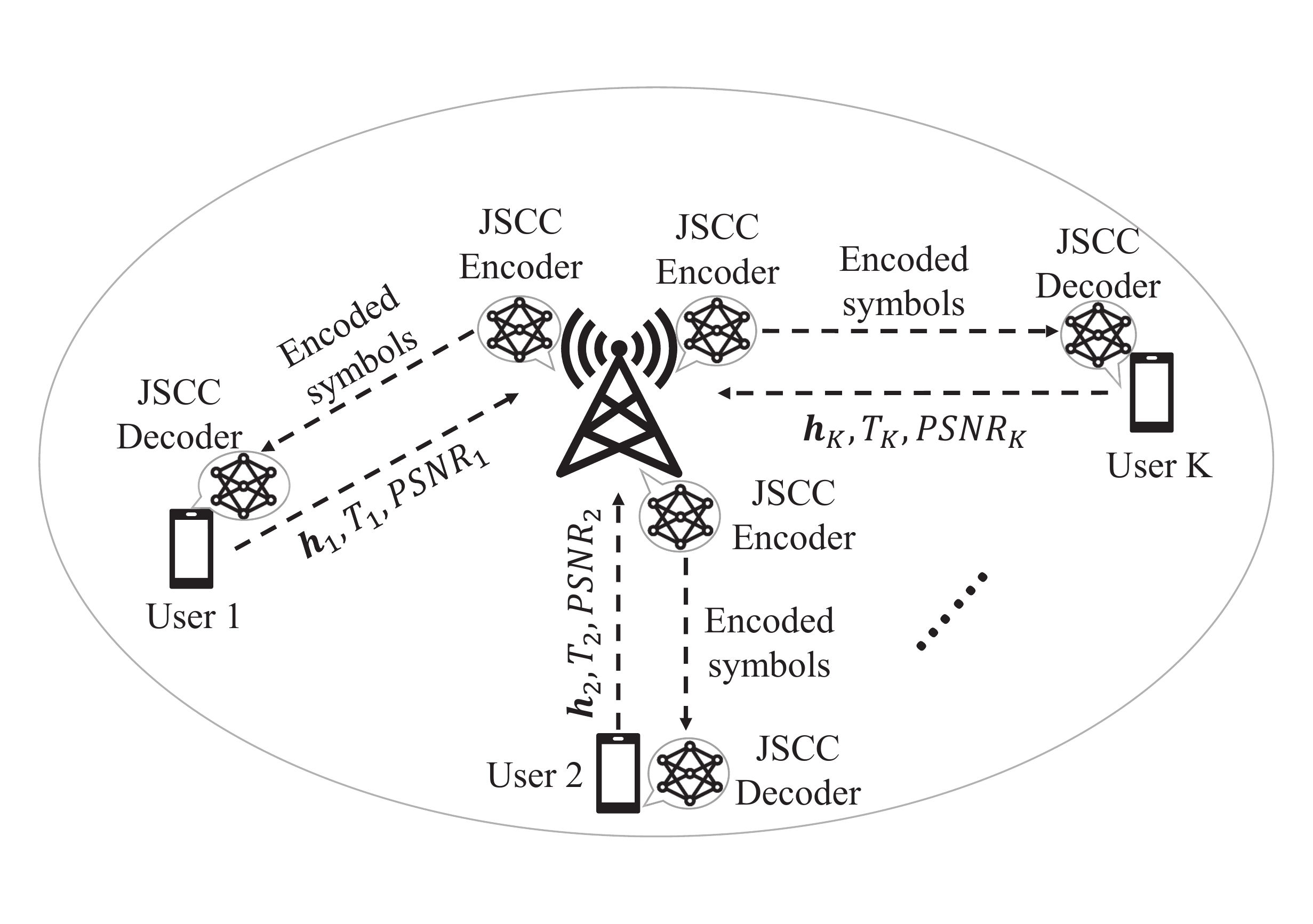}}
\caption{The structure of the JSCC based communication systems.}
\label{fig_structure}
\end{figure}

We consider a cellular network in which a base station (BS) transmits signal to a set $\mathcal{K}$ of $K$ users using JSCC networks as illustrated in Fig.~\ref{fig_structure}. Each user aims at accomplishing a specific task, such as image transmission, text transmission, etc. Without loss of generality, we assume all the users aim for image transmission and employ the deep learning based joint source-channel coding scheme proposed in \cite{jscc}. The encoder and decoder of the JSCC scheme are deployed at the BS and the users, respectively. We emphasize that the proposed resource allocation method can be applied for any other JSCC scheme for other tasks. We assume that different users have different task performance requirements, so they require different amounts of the transmitted data. We also assume the users are latency-sensitive with strict latency constraints. The considered downlink transmission operates as follows: Each user sends its performance constraint and delay constraint as well as channel state information (CSI) to the BS. According to the information received from all users, the BS performs resource allocation to maximize the capacity of the system.

\vspace{-0.5em}
\vspace{-0.5em}

\subsection{Transmitter and Receiver}

Denote that the raw image size is $D_{i}$, and the amount of actual transmitted data is $d_{k} = D_{i}o_{k}$, where $o_{k}$ denotes the compression ratio, which is defined as the proportion of transmitted symbols versus the total symbols of the original image. We assume that the compression ratio belongs to a set $\mathcal{N}$ of $N$ predefined compression ratios (CR), i.e., $o_{k} \in \{c_{1}, c_{2},..., c_{N}\}, \forall k \in \mathcal{K}$. Depending on users' requirements, the transmitter extracts feature vectors with different sizes.  After determining the proper compression ratio, BS adopts a proper well-trained JSCC network to encode the image and transmits the encoded information together with the index of the adopted network to receiver. The receiver then  applies the corresponding JSCC decoder to recover the received information according to the index.

We adopt the peak signal-to-noise ratio (PSNR) as the performance measure, given as:
\begin{equation}
    \mathrm{PSNR}=10 \log _{10} \frac{\mathrm{MAX}^{2}}{\mathrm{MSE}} \mathrm{(dB)},
\end{equation}
where $\mathrm{MSE}$ is the mean square error of the pixels between the original and the reconstructed images, and MAX denotes the maximum value of the image pixels. In our setting, $\mathrm{MAX}=2^8-1=255$. 

\begin{figure}[htbp]
\centerline{\includegraphics[width=6cm ,trim=20 10 20 10,clip]{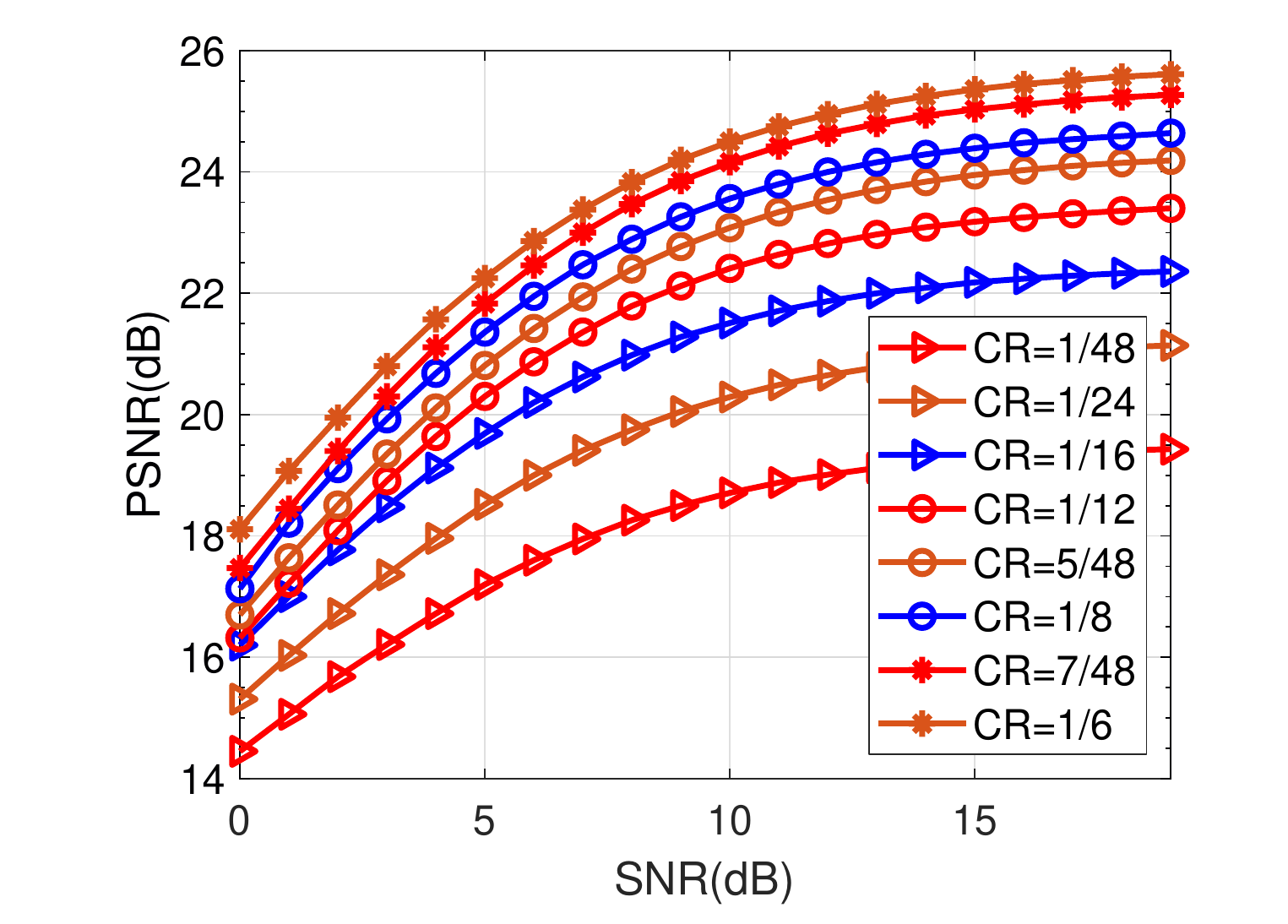}}
\caption{The PSNR of the reconstructed image vs. the SNR under different compression ratios.}
\label{fig_snr}
\end{figure}

The PSNR of the received image is determined by the compression ratio and the SNR of the received signal, i.e.
\begin{equation}
    PSNR_{k}=f\left(o_{k}, \gamma_{k}\right), \label{psnrk}
\end{equation}
where $\gamma_{k}$ is the SNR of user $k$. $f(*)$ are the functions that depend on the JSCC model. For example, Fig.~\ref{fig_snr} shows the PSNR versus SNR under different compression ratios of a JSCC scheme over an additive white Gaussian noise (AWGN) channel, and the simulation settings will be illustrated in Section \uppercase\expandafter{\romannumeral5}. We can obtain the approximate $f(*)$ experimentally by assuming a polynomial function. We note that our work can be easily applied to other JSCC schemes such as text \cite{deepsc} and speech \cite{speech} transmission, in which the performance measure is also determined by both the compression ratio and SNR.

\vspace{-0.5em}
\vspace{-0.5em}

\subsection{Transmission Model}
 We consider an orthogonal frequency division multiple access (OFDMA) system. The whole bandwidth $B$ with $S$ sub-channels are uniformly divided into a set $\mathcal{M}$ of $M$ orthogonal RBs, which can be allocated to the users, each RB contains $S_0 = \lfloor \frac{S}{M} \rfloor$ sub-channels. In each RB, the feature vectors are modulated with orthogonal frequency division multiplexing (OFDM). The allocation vector of user $k$ is $\boldsymbol{\alpha}_{k}=\left[\alpha_{k, 1}, \cdots, \alpha_{k, m}, \cdots, \alpha_{k, M}\right]$ , where $\alpha_{k, m} \in\{0,1\}$. $\alpha_{k, m}=1$ denotes that the $m \text {-th}$ RB is allocated to the $k \text {-th}$ user, and $\alpha_{k, m}=0$, otherwise. We assume that each RB can only be allocated to at most one user and each user can only occupy at most one RB, i.e.,
\begin{equation}
    \sum_{k=1}^{K} \alpha_{k, m} \leq 1, \forall m \in \mathcal{M} ; \sum_{m=1}^{M} \alpha_{k, m} \leq 1, \forall k \in \mathcal{K}.
\end{equation}

For each user, the channel gain between BS and user $k$ is denoted as $\bm{h} _{k} \in \mathcal{C} ^{S*1} $, the channel gain between the BS and user $k$ on sub-channel $s$ is $h_{k,s}$. We note that it takes less than 10ms to transmit an image in our setting, so we assume that the channel gain remains constant during the transmission of a image. Assume that RB $m$ is allocated to user $k$, i.e., $\alpha_{k,m}=1$, the received signal at the $s \in \{(m-1)*S_{0}+1, (m-1)*S_{0}+2,...,m*S_{0}\}$ sub-channel is $y_{k,s} = h_{k,s} \sqrt{p_{k,s}} x_{k,s} + n_{0}$, where $p_{k,s}$ is the power allocated to user $k$ on the sub-channel $s$, $x_{k,s}$ is the normalized transmitted symbol of user $k$ on the sub-channel $s$, $n_{0} \in \mathcal{CN}(0, N_{0})$ is the noise power per channel. Thus, the received SNR of user $k$ on the sub-channel $s$ is $\gamma_{k, s}=10 \lg \left(\frac{\left|h_{k, s}\right|^{2} p_{k, S}}{N_{0}}\right)$.

Inspired by the over-the-air computation in \cite{over-the-air}, different from the classic OFDM transmitter, we replace digital modulation (e.g., QAM) with analog one. More specifically, the extracted feature vector is divided into blocks, and each block is transmitted by an OFDM symbol with one signal over one sub-channel. We emphasize that the BS performs power allocation in a way such that the signals belonging to the same user have the same SNR, achieving amplitude alignment at the receiver. Given a fixed SNR of user $k$ as $\gamma_{k}$, the power allocated to user $k$ on sub-channel $s$ is $p_{k, s}=\frac{N_{0}}{\left|h_{k, s}\right|^{2}} 10^{\frac{\gamma_{k}}{10}}$. Summing over all the allocated sub-channels, the power allocated to user $k$ on $m$-th RB is
\begin{equation}
    p_{k,m}= \sum_{i=1}^{S_{0}} p_{k,(m-1) * S_{0}+i}, \forall k \in \mathcal{K}.  \label{pk}
\end{equation}
The received signals of user $k$ are then normalized by $\hat{x}_{k, s}=\frac{y_{k, s}}{h_{k, s} \sqrt{p_{k, s}}}$, by which we can view the OFDM channel as an AWGN channel with an SNR of $\gamma_{k}$, where $\gamma_{k}=10 \lg \left(\frac{\left|h_{k, s}\right|^{2} p_{k, s}}{N_{0}}\right), s \in \{(m-1)*S_{0}+1, (m-1)*S_{0}+2,...,m*S_{0}\}$ is equal in all the sub-channels allocated to user $k$.

\vspace{-0.5em}
\section{Problem Formulation}
\vspace{-0.5em}
In this section, we first define an utility function, which reflects whether the constraints of each user are met or not followed by the formulation of the considered resource allocation problem.

\vspace{-0.5em}
\subsection{User Utility Function}\label{AA}
\vspace{-0.5em}
If being allocated a RB for transmission, the transmission delay of an image to user $k$ is 
\begin{equation}
    t_{k}=\frac{D_{i} * o_{k}}{S_{0}} * T_{0}, \label{tk}
\end{equation}
where $T_{0}$ is the symbol duration of an OFDM symbol.

We denote the transmission delay constraint for user $k$ by $T_{k}$, $k=1, ..., K$. Besides, each user also has a performance constraint to guarantee the quality of the reconstructed images, i.e.,
\begin{equation}
    PSNR_{k} \geq \eta_{k},
\end{equation}
where $\eta_{k}$ is the minimum PSNR that user $k$ needs to satisfy.

The utility of user $k$, $U_{k}$, is set to 1 when latency and reconstruction quality requirements are all satisfied given allocated resources. The utility function can be expressed as
\begin{equation}
U_{k}=\left\{
\begin{aligned}
    & 1, \quad \text { if } t_{k} \leq T_{k}, PSNR_{k} \geq \eta_{k} \\
    & 0, \quad \text { otherwise }.
\end{aligned}
\right.
\end{equation}

\subsection{Problem Formulation}
In the considered system, our goal is to maximize the number of users that satisfy the transmission delay constraint and performance constraint in terms of optimizing the RB allocation, power allocation and compression ratio of each user. The optimization problem is formulated as follows:
\begin{subequations}
    \begin{align}
    \mathcal{P}1:\  & \max _{\alpha_{k, m}, o_{k}, p_{k,m}} \  \sum_{k=1}^{K} \  U_{k}, \\
    \text {s.t.} \ & \alpha_{k, m} \in\{0,1\}, \  \forall k \in \mathcal{K}, \ \forall m \in \mathcal{M}, \label{p1b} \\
    & \sum_{k} \alpha_{k, m} \leq 1, \  \forall m \in \mathcal{M}, \label{p1c} \\
    & \sum_{m} \alpha_{k, m} \leq 1, \  \forall k \in \mathcal{K}, \label{p1d} \\
    & \sum_{k=1}^{K} \sum_{m=1}^{M} \alpha_{k, m} p_{k,m} \leq P, \label{p1e} \\
    & o_{k} \in \{c_{1}, c_{2},..., c_{N}\}, \  \forall k \in \mathcal{K}. \label{p1f}
    \end{align}
\end{subequations}
where \eqref{p1b}, \eqref{p1c} and \eqref{p1d} are RB allocation constraints, which guarantee that each RB can only be allocated to at most one user and each user can only occupy at most one RB, \eqref{p1e} indicates that the total transmission power cannot exceed the power that BS can provide, and \eqref{p1f} limits the range of compression ratio. It should be noted that $\mathcal{P}1$ is a mixed integer non-linear problem, which is hard to solve. In the next section, we will solve it by decomposing it into two subproblems.

\vspace{-0.3em}
\section{Our solution}
\vspace{-0.3em}
In this section, we first decouple the original problem into two subproblems followed by the corresponding solutions. At last, the complexity of the proposed problem is investigated.

\subsection{Problem decomposition}
The main difficulty in solving $\mathcal{P}1$ is that the problem involves the discrete variables $\left \{ \alpha _{k, m}, o_{k}  \right \} $ and continuous variables $\left \{ p_{k, m}  \right \} $. However, by analyzing the problem, we can decompose it into two subproblems and then solve them separately.
It should be noted that our goal is to maximize the number of supported users in the system, and the main restriction is the limited power of the BS. We can first minimize the required power for each given user-RB pair $(k, m)$ by optimizing the compression ratio and power allocation $\left \{ o_{k}, p_{k, m}  \right \} $. Let $p_{k, m}$ denote the required BS power allocation that guarantees user $k$ allocated with $m$-th RB to accomplish its transmission task satisfying the performance and latency constraints.
The mathematical formulation for minimizing the required power allocation of user-RB pair $(k,m)$ can be expressed as
\begin{subequations}
    \begin{align}
    \mathcal{P}2:\  & \min _{o_{k}, p_{k, m}} \   p_{k, m}, \\
    \text {s.t.}\  & PSNR_{k} \geq \eta_{k}, \\
    & t_{k} \leq T_{k}, \\
    & o_{k} \in \{c_{1}, c_{2},..., c_{N}\}.
    \end{align}
\end{subequations}

After obtaining the minimum required power, denoted by $p_{k, m}^{*}$, by solving $\mathcal{P}2$ for each user-RB pair $(k, m)$, we then maximize the number of supported users by optimizing the user-RB paring. Thus, the original problem, i.e., $\mathcal{P}1$, can be reformulated as
\begin{subequations}
    \begin{align}
    \mathcal{P}3:\  & \max _{\alpha_{k, m}} \  \sum_{k=1}^{K} U_{k}, \\
    \text {s.t.}\  & \mathrm{\eqref{p1b}}, \mathrm{\eqref{p1c}}, {\rm and}\  \mathrm{\eqref{p1d}}, \\
    & \sum_{k=1}^{K} \sum_{m=1}^{M} \alpha_{k, m} p_{k, m}^{*} \leq P. \label{p3c}
    \end{align}
\end{subequations}
where \eqref{p3c} guarantees the power constraint. 

Thus, we decouple the $\mathcal{P}1$ into $\mathcal{P}2$ and $\mathcal{P}3$. By evaluating the inner relations between the original problem and two subproblems, we have the following theorem.

\textit{Theorem 1:} The combination of optimal solutions of $\mathcal{P}2$ and $\mathcal{P}3$ is equal to the optimal solution of $\mathcal{P}1$.

\textit{Proof:} Please see Appendix A.

\subsection{Power Minimization}
In this subsection, we discuss how to obtain the minimum power for each user-RB pair $(k, m)$. According to \eqref{psnrk} and \eqref{tk}, we find that $PSNR_{k}$ is a function of $o_{k}$ and $p_{k, m}$ while $t_{k}$ is only determined by $o_{k}$. With given latency constraint, we can obtain the maximum compression ratio according to \eqref{tk}, which can be denoted as
\begin{equation}
    o_{max} = \frac{T_{k} S_{0}}{T_{0} D_{i}}. \label{ok_solution}
\end{equation}

According to Fig.\ref{fig_snr}, when $PSNR_{k}$ is fixed, the $\gamma_{k}$ is approximately inversely proportional to the compression ratio $o_{k}$, i.e., the larger the $o_{k}$ is, the smaller the $\gamma_{k}$ will be required, and the less the power $p_{k, m}$ will be required. As a result, in order to minimize the power allocation of user-RB pair $(k,m)$, we need to search the maximum $o_{k}$ that satisfies $o_{k} \leq o_{max}$ within the compression ratio sets $\mathcal{N}$ as the selected compression ratio $o_{k}^{*}$, i.e., $o_{k}^{*}=\max _{o_{k} \in \mathcal{N}, o_{k} \leq o_{\max }} o_{k}$.

Recall that we use a polynomial function to approximate \eqref{psnrk} according to Fig.\ref{fig_snr}. When the compression ratio is determined, we can obtain the minimum SNR that satisfies the PSNR constraint using bisection method by solving the problem
\begin{equation}
    f(o_{k}^{*}, \gamma_{k}) - \eta_{k} = 0, \label{pk_psnr}
\end{equation}
when the minimum SNR $\gamma_{k}^{*}$ is searched, the minimum power cost of user-RB pair $(k, m)$ can be obtained by \eqref{pk}.

\subsection{System Capacity Maximization}
We note that the goal of $\mathcal{P}3$ is to maximize the number of supported users under certain constraints on power resources. Thus, we assume that there are $J(0 \leq J \leq K)$ users that can be supported, and we then calculate the minimum power that guarantees the completion of tasks for these $J$ users, which is denoted as $P_{J}$. We can get the following optimizing problem
\begin{subequations}
    \begin{align}
    \mathcal{P}4:\  & \min _{\alpha_{k, m}} \  P_{J}=\sum_{k=1}^{K} \sum_{m=1}^{M} \alpha_{k, m} p_{k, m}^{*} \\
    \text {s.t.}\  & \alpha_{k, m} \in\{0,1\},  \  \forall k \in \mathcal{K}, \ \forall m \in \mathcal{M}, \\
    & \sum_{k} \alpha_{k, m} \leq 1, \forall m \in \mathcal{M}, \label{p4c} \\
    & \sum_{m} \alpha_{k, m} \leq 1, \forall k \in \mathcal{K}, \label{p4d} \\
    & \sum_{k=1}^{K} \sum_{m=1}^{M} \alpha_{k, m}=J. \label{p4e}
    \end{align}
\end{subequations}

Denote the optimal value of $\mathcal{P}4$ as $P_{J}^{*}$. After that, we can compare the optimal value $P_{J}^{*}$ with the total power $P$ that the BS can provide and test the feasibility of the solution. If $P \geq P_{J}^{*}$, the total power is sufficient so that all $J$ users can be supported. Otherwise, the solution is infeasible. Therefore, we can get the optimal solution of $\mathcal{P}3$ via exhaustively searching $J$ from $K$ to 0 until the optimal value of $\mathcal{P}4$ is infeasible.

In order to solve $\mathcal{P}4$, we relax integer variables, $\alpha_{k, m}$, into continuous variables, and we can obtain the following problem
\begin{subequations}
    \begin{align}
    \mathcal{P}5:\  & \min _{\alpha_{k, m}} \  P_{J}=\sum_{k=1}^{K} \sum_{m=1}^{M} \alpha_{k, m} p_{k, m}^{*} \\
    \text {s.t.}\  & 0 \leq \alpha_{k, m} \leq 1 \\
    & \mathrm{\eqref{p4c}}, \mathrm{\eqref{p4d}}, {\rm and} \  \mathrm{\eqref{p4e}}.
    \end{align}
\end{subequations}
This is a typical linear programming problem and it can be solved by 
Karmarkars's algorithm \cite{karmarkar}. Although the integer variables have been relaxed, it has been proven in \cite{d2d} that the optimal value in $\mathcal{P}5$, denoted by $\alpha_{k, m}^{*}$ is either 0 or 1, which means $\mathcal{P}5$ and $\mathcal{P}4$ have the same optimal solution.

\begin{algorithm}[H]
    \caption{The Optimal Algorithm to $\mathcal{P}1$}
    \label{alg:algorithm1}
    \begin{algorithmic}[1] 
        \STATE {Initialize $p_{k, m}^{*}$ by solving $\mathcal{P}2$, $\forall k \in \mathcal{K}, \forall m \in \mathcal{M}$.}
        \STATE {Set $J_{u} = K$, $J_{l} = 0$ as the upper bound and lower bound.}
        \STATE {Calculate $P_{J_{u}}^{*}$ by solving $\mathcal{P}5$ with $J=J_{u}$.}
        \IF {$P \geq P_{J_{u}}^{*}$}
            \STATE {\textbf{return} $J_{u}$.}
        \ELSE 
            \WHILE {$J_{u} - J_{l} > 1$}
                \STATE {$J = \lfloor (J_{l} + J_{u}) / 2 \rfloor$.}
                \STATE {Calculate $P_{J}^{*}$ by solving $\mathcal{P}5$.}
                \IF{$P_{J}^{*} > P$}
                    \STATE{$J_{u} = J$.}
                \ELSE
                    \STATE{$J_{l} = J$.}
                \ENDIF
            \ENDWHILE
            \STATE{\textbf{return} $J_{l}$.}
        \ENDIF
    \end{algorithmic} 
\end{algorithm}

Note that it may be time consuming to search $J$ with $\mathcal{P}5$ from $K$ to 0. We adopt bisection search method instead to search the optimal value $J^{*}$, which would significantly reduce the iteration number. We denote $J_{u}$ and $J_{l}$ as the upper bound and lower bound of the searching space. The pseudo-code is presented in Algorithm 1.

\vspace{-0em}
\subsection{Complexity Analysis}
\vspace{-0em}
In $\mathcal{P}2$, we need to compute the minimum required power for all $K*M$ potential user-RB pairs, the number of required iterations using bisection method is $\mathcal{O}(\log(L_{0} / e))$, where $L_{0}$ is the initial interval length of the search region, and $e$ is the required precision. In $\mathcal{P}5$, the required iteration number of the bisection searching is $\mathcal{O}(\log(J_{u} - J_{l}))$. For each given $J$, $\mathcal{P}5$ is solved by Karmarkar's algorithm, which has the complexity of $\mathcal{O}((KM)^{3.5}L\log{L}\log{\log{L}})$, where $L$ is the length of bits required to enter into the computer. Thus, the computational complexity of the whole algorithm is $\mathcal{O}(KM(\log(L_{0}/e)) + (\log(J_{u} - J_{l}))(KM)^{3.5}L\log{L}\log{\log{L}})$, which is polynomial.

\vspace{-1em}
\section{Simulation Results}
\vspace{-0.3em}
In this section, we present the simulation results to demonstrate the performance enhancement of the proposed resource allocation method. The simulation settings are as follows unless otherwise stated. In our simulation, the BS has a radius of 500 m and 30 users are randomly located within this range. And we assume that $M = 30$. The maximum power that the BS can provide is 1 W. The adopted pathloss is 128.1+37.6lg[d(km)] dB and the shadowing factor follows a uniform distribution from 0 to 10 dB. The small-scale fading follows Rayleigh fading with zero mean and unit variance. The total bandwidth is $B=3$ MHz, the number of sub-channels is 100 and the bandwidth of a sub-channel is $30$ kHz, the OFDM symbol duration is 33.3 $\mu$s. The noise power is $\sigma^{2} = -114$ dBm. The delay tolerance of each user, i.e., $T_{k}$, is uniformly generated from 4 to 6 ms. For the performance requirement, the PSNR requirement of the user follows the uniform distribution within $\eta_{k} \in [20, 25]$ dB. We evaluate our system on CIFAR10 dataset \cite{cifar}, which consists of 60000 images with $32 \times 32$ resolution. We adopted the JSCC model in \cite{jscc} with a fixed training SNR of 10dB, and the test SNR varies from 0 to 20dB. The optional compression ratio set is $\{1/6, 7/48, 1/8, 5/48, 1/12, 1/16, 1/24, 1/48\}$.

For comparison, we consider three benchmarks: a) the first one solves the second subproblem with Hungarian algorithm \cite{hungarian}, which is equivalent to directly solving $\mathcal{P}4$ with $J=K$. By this way, we can obtain the minimum required power of each user. After that, we choose the user with greedy algorithm according to the required power. More specifically, the user which required less power will be choose first until the sum power of the selected users exceeds the BS power, named ``Hungarian Based". b) the minimum power is obtained according to $\mathcal{P}2$ while the user selects RB randomly. Besides, users will also be chosen greedily according to the power cost, named ``Random Pairing". c) the power is allocated uniformly and the user selects RB randomly, named ``Uniform Power".

\begin{figure}[htbp]
\centerline{\includegraphics[width=6cm ,trim=20 5 20 10,clip]{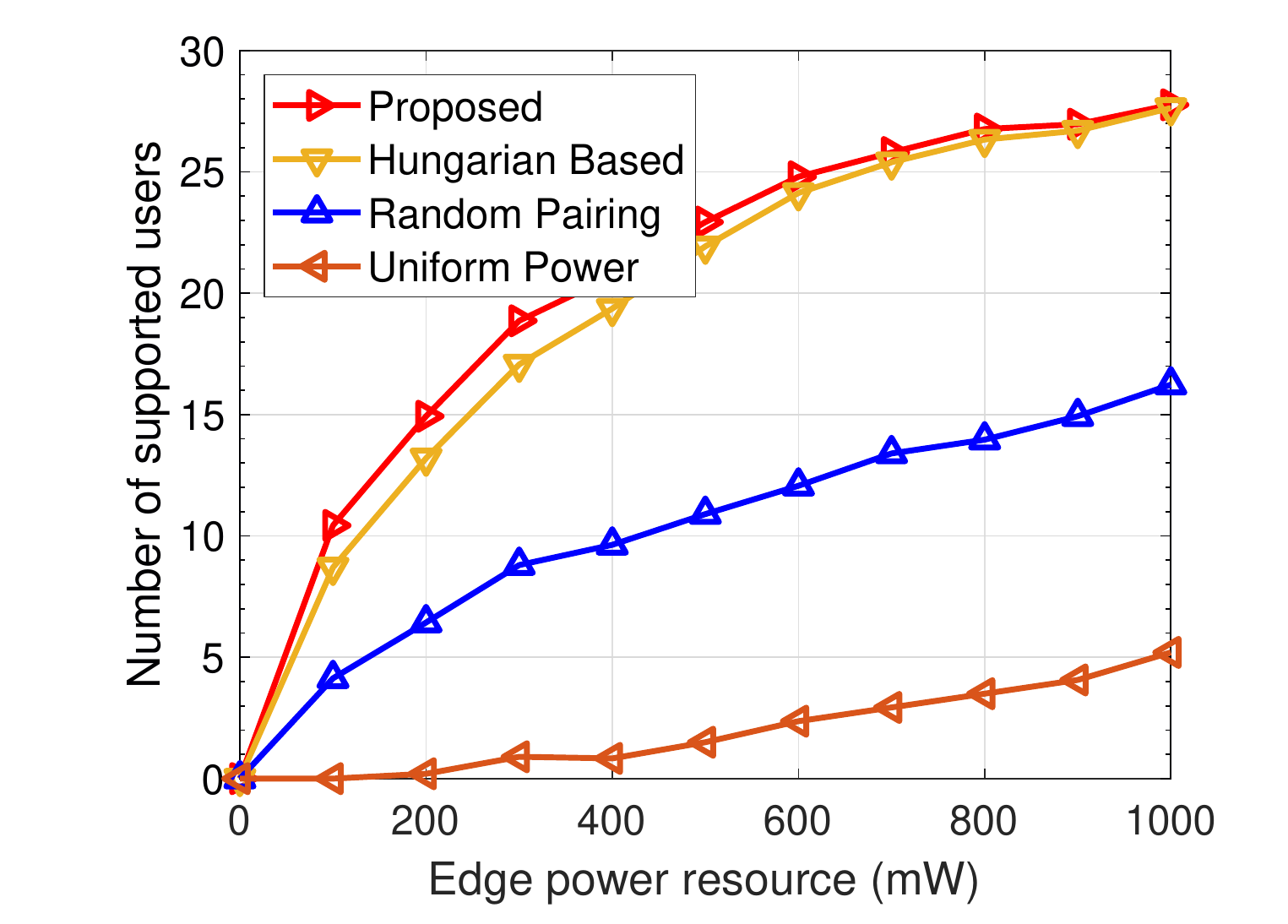}}
\caption{The number of the supported users vs. the BS power.}
\label{fig_pk}
\end{figure}

Fig.~\ref{fig_pk} shows the number of the supported users versus the BS power. It can be observed that the number of the supported users of both the proposed method and benchmarks increase with the BS power. Besides, the proposed method always achieves the best performance compared with all the benchmarks. It can be observed that the performance gap between the proposed method and the Hungarian based method is small, this is because Hungarian based method is optimal when power is adequate to support all users, but it becomes sub-optimal when the power is limited. This result reveals that the proposed method can effectively exploit the limited BS power.

\vspace{-1em}
\vspace{-0.5em}

\begin{figure}[htbp]
\centerline{\includegraphics[width=6cm ,trim=20 5 20 10,clip]{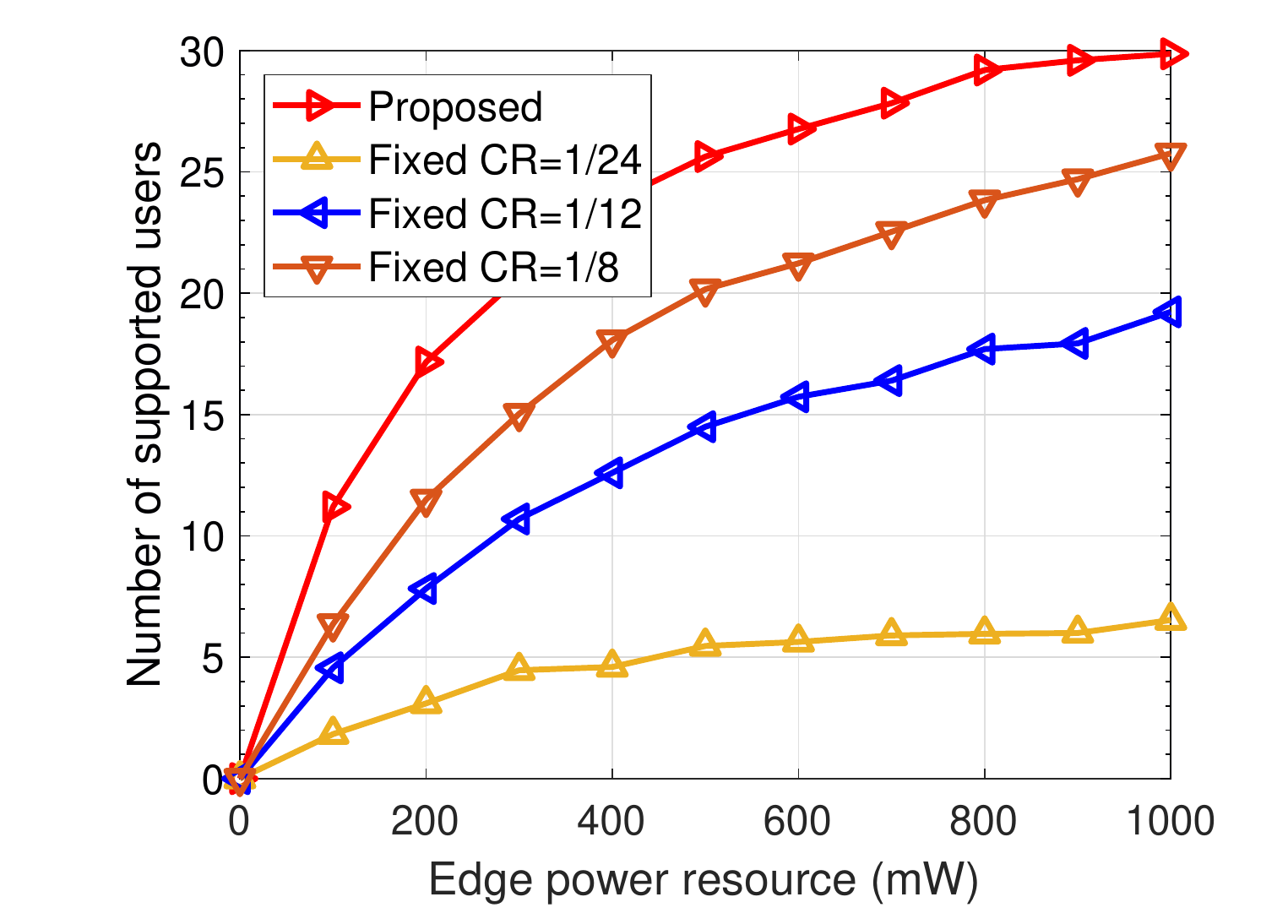}}
\caption{The number of the supported users vs. the BS power.}
\label{fig_com}
\end{figure}

To exploit the effectiveness of the adaptive compression scheme, we compare the proposed method with fixed compression ratio (CR) methods. We note that the resource allocation schemes of these methods are the same as the proposed method. Fig.~\ref{fig_com} shows the number of the supported users versus the BS power. From the figure, we can observe that fixed compression methods show degraded performance compared to the proposed method. The reason is that improper fixed compression may lead to higher SNR requirement, which would lead to additional power cost. Besides, the method with a fixed compression ratio of 1/24 shows significant performance degradation than others. This is because it transmits few symbols, thus it could not satisfy the performance constraint of most users even allocate high power to it.

\vspace{-1em}
\vspace{-0.5em}

\begin{figure}[htbp]
\centerline{\includegraphics[width=6cm ,trim=20 5 20 10,clip]{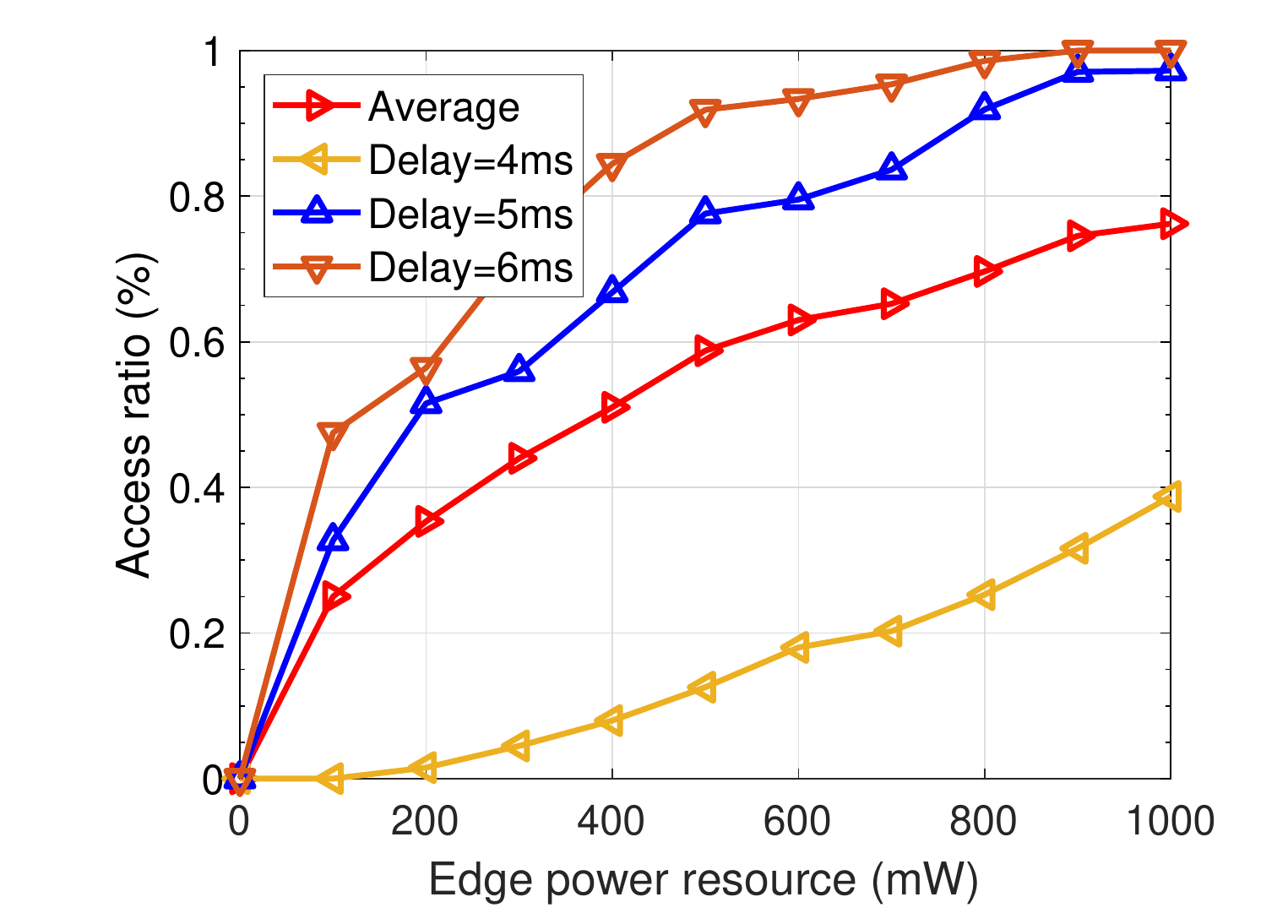}}
\caption{Access ratio vs. the BS power.}
\label{fig_delay}
\end{figure}

Fig.~\ref{fig_delay} illustrates the access ratio, which is defined as the proportion of users that can be supported versus the total number of users, versus delay constraints in the case of same PSNR constraint. In this case, all users have the same PSNR constraint, which is set as 24, and their delay requirements are randomly selected in three optional value, i.e. 4ms, 5ms, and 6ms.  As mentioned before, when the PSNR constraint remains unchanged, the smaller the required delay is, the smaller the compression ratio is, thus, the greater the required transmission power will be. Furthermore, we can observe that the access ratio of users with 4ms delay remains low. The reason is that, these users need quite large power to satisfy the PSNR constraint while the BS is not able to provide that much power. 

\vspace{-1em}
\vspace{-0.5em}

\begin{figure}[htbp]
\centerline{\includegraphics[width=6cm ,trim=20 5 20 10,clip]{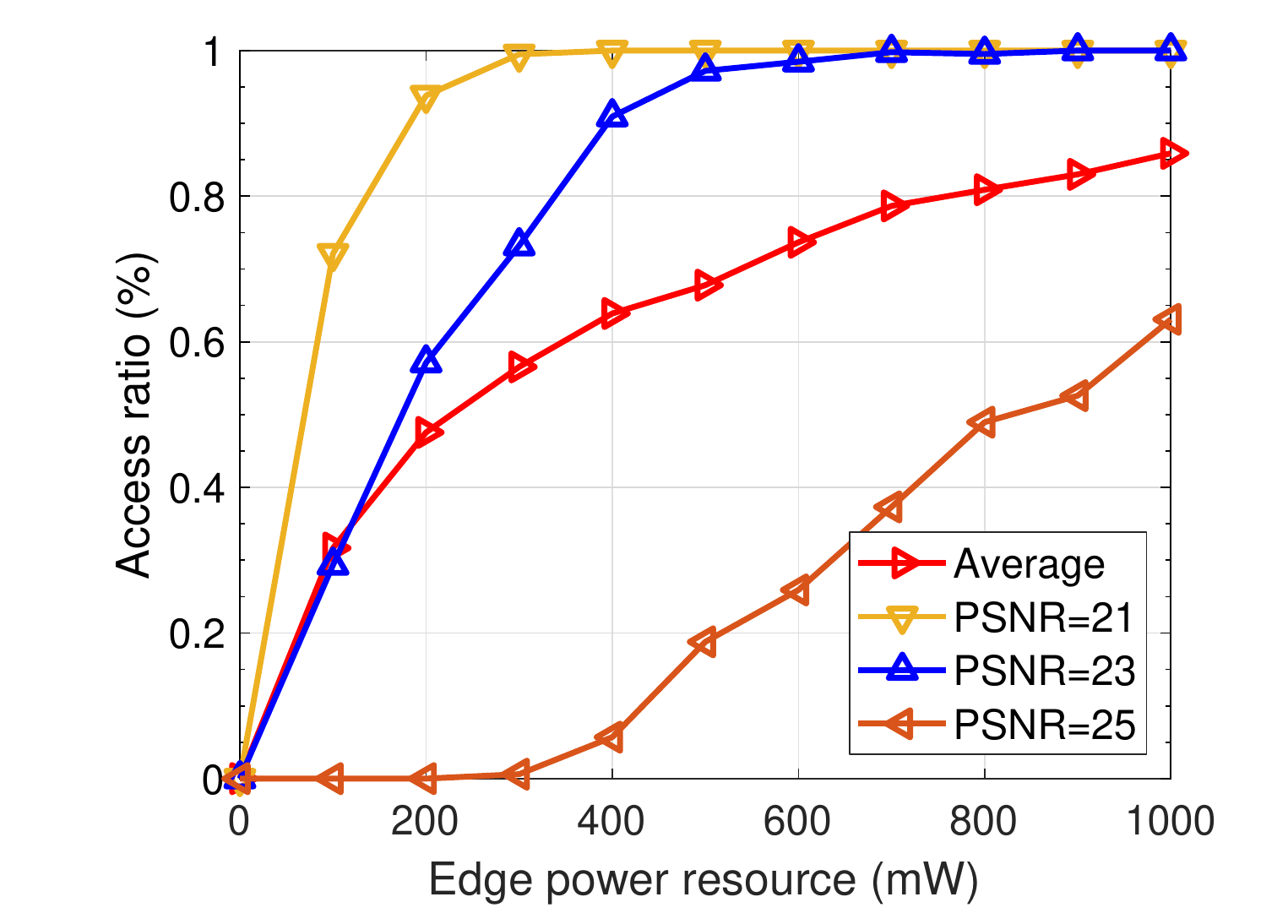}}
\caption{Access ratio vs. the BS power.}
\label{fig_psnr}
\end{figure}

Fig.~\ref{fig_psnr} shows the access ratio versus PSNR constraints in the case of same latency demand. In this case, all users have the same latency demand, which is set as 5ms, and their PSNR requirements are randomly selected in three optional value, i.e. 21, 23, and 25. It is obvious that when the latency requirements remain the same, the maximum compression ratio is fixed, higher PSNR requirement means that higher SNR is required, which leads to more power demand. Similarly, those users with high PSNR needs quite large power beyond what the base station can provide.

\vspace{-1em}
\vspace{-0.5em}

\vspace{-1em}
\section{Conclusion}
\vspace{-0.5em}
In this paper, we studied the resource allocation of JSCC networks which aims at maximizing the system capacity, i.e., the number of supported users, of a OFDM communication system.  We formulated this considered problem as an mixed integer non-linear optimization problem. We then decoupled it into two subproblems and proved that the optimal solutions of the two subproblems compose the optimal solution of the original problem. The first subproblem was solved by searching the maximum compression ratio within the feasible region. The second subproblem was solved by bisection search as well as Karmarkar's algorithm. Extensive simulations have been conducted to demonstrate the performance of the proposed method. Compared with benchmarks, the proposed algorithm could support more users when the power of BS is limited. We also showed that adaptive compression ratio instead of fixed scheme remarkably improves the system capacity. 

\vspace{-0.5em}
\section*{Appendix A}
\vspace{-0.5em}
We note that $\mathcal{P}1$ and $\mathcal{P}3$ have the same objective function, so we can prove that the optimal value of $\mathcal{P}1$ is also the optimal value of $\mathcal{P}3$, and vice versa. 


Denote $\{\alpha_{k, m}^{*, 1}, p_{k, m}^{*, 1}\}$ as the optimal solution of $\mathcal{P}1$ and $J^{*, 1} = \sum_{k=1}^{K} \  U_{k}$ as the optimal value. Let $p_{k, m}^{*, 2}$ denote the optimal value of $\mathcal{P}2$. Let $\{\alpha_{k, m}^{*, 3}\}$ denote the optimal solution of $\mathcal{P}3$ and $J^{*, 3} = \sum_{k=1}^{K} \  U_{k}$ denotes the optimal value. Since $\mathcal{P}2$ optimize the required power given specific user-RB pair $(k, m)$ , $p_{k, m}^{*, 2}$ must be no more than $p_{k, m}^{*, 1}$, i.e., $p_{k, m}^{*, 2} \leq p_{k, m}^{*, 1}$. Thus, we have
\begin{equation}
    \sum_{k=1}^{K} \sum_{m=1}^{M} \alpha_{k, m}^{*, 1} p_{k, m}^{*, 2} \leq \sum_{k=1}^{K} \sum_{m=1}^{M} \alpha_{k, m}^{*, 1} p_{k, m}^{*, 1} \leq P
\end{equation}
which means that $\alpha_{k, m}^{*, 1}$ satisfy constraint \eqref{p3c}. Besides, $\alpha_{k, m}^{*, 1}$ also satisfy other constraints of $\mathcal{P}3$. As a result, it is a feasible solution to $\mathcal{P}3$, so $J^{*, 1}$ is no more than $J^{*, 3}$, i.e., $J^{*, 1} \leq J^{*, 3}$.

Similarly, the optimal solution of $\mathcal{P}2$, $p_{k, m}^{*, 2}$, satisfies constraint \eqref{p1f}. The optimal solution of $\mathcal{P}3$, $\{\alpha_{k, m}^{*, 3}\}$, satisfies constraints \eqref{p1b}, \eqref{p1c} and \eqref{p1d}. Since $\{\alpha_{k, m}^{*, 3}, p_{k, m}^{*, 2} \}$ satisfy \eqref{p3c}, it also satisfy \eqref{p1e}. As a result, $\{\alpha_{k, m}^{*, 3}, p_{k, m}^{*, 2}, o_{k}^{*, 2}\}$ is a feasible solution to $\mathcal{P}1$, which means that $J^{*, 1}$ is no less than $J^{*, 3}$, i.e., $J^{*, 1} \geq J^{*, 3}$. Hence, the optimal value of $\mathcal{P}1$ is equal to the optimal value of $\mathcal{P}3$.

\vspace{10pt}


\end{document}